\documentclass[manuscript]{aastex}

\usepackage{natbib}
\shorttitle{}
\shortauthors{}

\begin{document}

\title{Four dual AGN candidates observed with the VLBA}

\author{K.\'E. Gab\'anyi\altaffilmark{1}} 
\affil{F\"OMI, Satellite Geodetic Observatory, P.O. Box 585, 1592 Budapest, Hungary}
\email{gabanyik@sgo.fomi.hu}
\author{T. An}
\affil{Shanghai Astronomical Observatory, Chinese Academy of Sciences, 80 Nandan Road, 200030 Shanghai, P.R. China}
\email{antao@shao.ac.cn} %%%%%NEW ADDITION
\author{S. Frey}
\affil{F\"OMI, Satellite Geodetic Observatory, P.O. Box 585, 1592 Budapest, Hungary}
\author{S. Komossa}
\affil{Max-Planck-Institut f\"ur Radioastronomie, Auf dem H\"ugel 69, 53121 Bonn, Germany}
\author{Z. Paragi}
\affil{Joint Institute for VLBI ERIC, Postbus 2, 7990 AA Dwingeloo, the Netherlands}
\author{X.-Y. Hong and Z.-Q. Shen}
\affil{Shanghai Astronomical Observatory, Chinese Academy of Sciences, 80 Nandan Road, 200030 Shanghai, P.R. China}

\altaffiltext{1}{Konkoly Observatory, Research Centre for Astronomy and Earth Sciences, Hungarian Academy of Sciences, Budapest, Hungary}

\begin{abstract}
According to hierarchical structure formation models, merging galaxies are expected to be seen in different stages of their coalescence. However, currently there are no straightforward observational methods neither to select nor to confirm a large number of dual active galactic nuclei (AGN) candidates. Most attempts involve the better understanding of double-peaked narrow emission line sources, to distinguish the objects where the emission lines originate from narrow-line kinematics or jet-driven outflows from those which might harbour dual AGN. We observed four such candidate sources with the Very Long Baseline Array (VLBA) at 1.5\,GHz with $\sim$10 milli-arcsecond angular resolution where spectral profiles of AGN optical emission suggested the existence of dual AGN. In SDSS J210449.13$-$000919.1 and SDSS J23044.82$-$093345.3, the radio structures are aligned with the optical emission features, thus the double-peaked emission lines might be the results of jet-driven outflows. In the third detected source SDSS J115523.74+150756.9, the radio structure is less extended and oriented nearly perpendicular to the position angle derived from optical spectroscopy. The fourth source remained undetected with the VLBA but it has been imaged with the Very Large Array at arcsec resolution a few months before our observations, suggesting the existence of extended radio structure. In none of the four sources did we detect two radio-emitting cores, a convincing signature of duality.
\end{abstract}

%% Keywords should appear after the \end{abstract} command. The uncommented
%% example has been keyed in ApJ style. See the instructions to authors
%% for the journal to which you are submitting your paper to determine
%% what keyword punctuation is appropriate.
%% MAX 6

\section{Introduction}

Currently it is widely accepted that most of the massive galaxies harbour supermassive black holes (SMBHs) in their centres \citep{central_SMBH}. In hierarchical structure formation models \citep{strucform_CDM1, strucform_CDM2}, interactions and mergers between galaxies play an important role in their evolution and consequently in the growth of their central SMBHs \citep[e.g.,][]{SMBH_galaxy_evo}. It is expected that a particular phase in the merging process, systems with dual SMBHs must be observed in the Universe. Following the definition of \cite{dual_binary}, we refer to systems consisting of dual SMBHs (or dual AGN), where the separation of the two SMBHs are much larger than their influence radii. In contrast, binary SMBHs are those systems where the separation is smaller than the influence radius. In the literature, kpc-scale separation systems are referred to as dual AGN, and pc-scale separation systems as binary AGN \citep[e.g.,][]{colpi_review,volonteri_review,Muller-Sanchez}

The galaxy merging process can cause enhanced accretion onto the central SMBHs and thus initiate activity. One or both of the dual SMBHs may be active or re-activated. Simulations \citep[e.g.,][]{vanWassenhove} suggest that simultaneous activity is mostly expected at the late phases of mergers, at or below 10 kpc-scale separations. Additionally, recent work of \cite{radioselected_binaryAGN2} indicates that mergers trigger and tend to synchronize activity at separations of few kpc. Therefore dual active galactic nuclei (AGN) are expected to be observed, but they are not easily resolvable with the current observing facilities. The number of convincing dual AGN systems, mostly detected by X-ray and radio observations \citep[e.g.,][]{confirmed_dual1, beswick_dual, confirmed_dual2,confirmed_dual4,confirmed_dual5,confirmed_dual6,confirmed_dual7,confirmed_dual8,confirmed_dual9,confirmed_dual10, Comerford_Xray, Muller-Sanchez}, are relatively few compared to the theoretically predicted abundance \citep[see also,][]{Komossa_Zensus_review}. While very long baseline interferometric (VLBI) radio observations currently provide the highest spatial resolution, thus they would be a promising tool for the detection of dual AGN\footnote{In fact thanks to its superior resolution, the first binary AGN was discovered by VLBI technique \citep{confirmed_dual3}.}, only about $10$\,\% of AGN are radio-loud, therefore dual radio emitting active nuclei would be quite rare. Moreover, when trying to confirm the existence of dual AGN in candidate sources, the non-detection in radio does not immediately falsify the dual AGN hypothesis because radio-quiet nuclei may as well be present. \cite{B-S_VLBI} searched the archival VLBI observations of 3114 radio-emitting AGN for sources containing compact double sources. Only one source (B3 0402$+$379) was found with double nucleus with separation of about $7$\,pc, which was already known from the observations of \cite{confirmed_dual3}.

On the other hand, not just the confirmation of dual AGN, but the selection of candidate sources is not straightforward. During the last decade, it was proposed that double-peaked narrow emission lines might be used to select possible dual AGN sources, where the two sets of narrow lines originate in the two distinct narrow-line regions (NLR) of the two AGN \citep[e.g.][]{double_line1, wang2009, double_line2}. However, it was also shown that this spectral behaviour can be explained by other effects occurring in single AGN. According to \citet{Heckman1981, Heckman1984}, double-peaked narrow emission lines can arise due to peculiar kinematics and jet--cloud interaction in a single NLR. More recently \cite{double_line2} discussed several mechanisms for creating double-peaked emission lines in single AGN, including a rotating disk-like NLR, and, in particular, the possibility of a {\em single} AGN illuminating the interstellar media of {\em two} galaxies in a merger. Further, \cite{non_dual} pointed out that blobby NLRs and extinction effects can produce multi-peaked narrow emission lines \citep[see also][]{Crenshaw2}. To select the real dual AGN from candidate sources, \cite{Tingay_VLBI} conducted VLBI observations of a sample of AGN with double-peaked optical emission lines. Among the observed 11 sources, they have found no evidence for double radio cores. Our group also conducted VLBI search for double radio cores from optically-selected dual AGN candidates. In the first of a systematic study of dual AGN candidates, we performed VLBI observations of 3C 316, which is the most radio-loud source in the double-peaked narrow-line AGN sample \citep{Smith2010}, in order to confirm its duality. The VLBI image detected a series of discrete compact knots, however, none of which could be compact enough to be identified as an AGN core. The most possible explanation of the observed radio structure is that it is the radio jet of a single radio-loud AGN in 3C 316. These works thus do not support the idea that double-peaked narrow emission lines are good tracer for dual AGN (although investigations of larger samples are needed to obtain firm statistical result).

\cite{Comerford_sample} proposed additional selection critera to narrow down the list of dual AGN candidates
from a group of sources showing double-peaked narrow lines. Using long-slit spectroscopy of $81$ double-peaked narrow-line AGN ($0.03\lesssim z \lesssim 0.3$), they tried to separate the sources where the double-peaked lines are the result of AGN outflows or gas kinematics from those which can be compelling dual AGN candidates. \cite{Comerford_sample} selected their targets from the catalogs of \cite{wang2009}, \cite{Liu2010}, and \cite{Smith2010}. They identified double-peaked active galaxies in the Sloan Digital Sky Survey (SDSS) spectral database using the [\ion{O}{3}] $\lambda5007$ emission lines. \cite{Comerford_sample} proposed that sources with spatially compact emission components may be preferentially produced by dual AGN. Moreover, they concluded that in more than one-third of their sample the double emission features are aligned with the major axis of the host galaxy, which is more than twice the expected amount assuming uniform distribution of the position angles. They proposed that this was also an indication of dual AGN in those objects. In summary, \cite{Comerford_sample} concluded that from the $81$ observed sources $17$ are promising dual AGN candidates with angular separation less than $1\arcsec$. In $14$ sources, the emission features are aligned with the major axis of the host galaxies, in $3$ sources the measured spatial offsets and position angles of the emission features are consistent with the separations and position angles of stellar components detected in adaptive optics images. 

Among these $17$ dual AGN candidates, seven have been detected in the Very Large Array (VLA) Faint Images of the Radio Sky at Twenty-centimeters (FIRST) survey \citep{first} at $1.4$\,GHz with flux densities between $0.8$\,mJy and $9.1$\,mJy. We selected the four sources, SDSS J102325.57$+$324348.4 (hereafter J1023+3243), SDSS J115523.74$+$150756.9 (hereafter J1155+1507), SDSS J210449.13$-$000919.1 (hereafter J2104$-$0009), and SDSS J23044.82$-$093345.3 (hereafter J2304$-$0933) which have integrated flux densities $\gtrsim 2$\,mJy in the FIRST survey, and observed them with the Very Long Baseline Array (VLBA) at $1.5$\,GHz. If two compact radio features can be detected at the separations and position angles derived by \cite{Comerford_sample} from the optical observations, that would be a clear indication of the existence of dual AGN in the sources. One compact AGN and extended jet-like feature aligned with the optical features may be indicative of jet-driven AGN outflow. However, the detection of only one or no compact radio emission does not exlcude the possibility of dual AGN in the galaxy, since the majority of AGN are not radio emitters.

Throughout the paper we assume a standard flat $\Lambda$CDM cosmological model with Hubble constant $H_0=70\mathrm{\,km\, s}^{-1}\mathrm{Mpc}^{-1}$, and density parameters $\Omega_\mathrm{m}=0.3$, and $\Omega_\Lambda=0.7$. The linear scales and luminosities were obtained with the online calculator of \cite{cosmocalc}.

\section{Observations and data reduction}

The ten 25-m diameter radio telescopes of the U.S. National Radio Astronomy Observatory (NRAO) VLBA were used for observing the four target sources at the central frequency of 1.5~GHz in phase-reference mode \citep[e.g.,][]{phase-ref}. Two sources (J1023+3243 and J1155+1507) were observed for 6~h on 2013 March 2 (experiment BA103A). Another 6-h experiment targeting J2104$-$0009 and J2304$-$0933 was performed on 2013 June 15 (experiment BA103B). The data were taken with 2~Gbit~s$^{-1}$ recording rate, leading to the total bandwidth of 256~MHz in both left and right circular polarizations using 2-bit sampling, in 8 separate intermediate-frequency (IF) channels, each with 32 spectral points. Phase-referencing was performed by periodically nodding the radio telescopes between the targets and the respective nearby bright and compact calibrator source. Nearly 3.5~min in each of the $\sim$4.5-min calibrator--target cycles were spent on the targets. The delay, delay rate, and phase solutions derived for the phase-reference calibrators (J1021+3437, J1157+1638, J2105+0033, and J2303$-$1002) could later be interpolated and applied for the respective target-source data, to improve the sensitivity of the observations of these weaker objects. The target--reference angular separations were $2\fdg05$ (between J1023+3243 and J1021+3437), $1\fdg61$ (between J1155+1507 and J1157+1638), $0\fdg72$ (between J2104$-$0009 and J2105+0033), and $0\fdg51$ (between J2304$-$0933 and J2303$-$1002). The total on-source time spent on each target was 2~h, the expected image thermal noise was $\sim$20~$\mu$Jy\,beam$^{-1}$. In addition, the strong fringe-finder sources 4C\,39.25 and 3C\,454.3 were occasionally observed in experiments BA103A and B, respectively.

The NRAO Astronomical Image Processing System\footnote{\url{http://www.aips.nrao.edu}} ({\sc AIPS}) was used for the data calibration \citep[e.g.,][]{data_reduc}. We followed the standard procedures applicable to the calibration of continuum VLBA data. This involved corrections for the dispersive ionospheric delay using maps of the total electron content derived from global navigation satellite systems measurements, for the accurately measured Earth orientation parameters, and for digital sampling effects. The visibility amplitudes were calibrated using system temperatures and antenna gains measured at the telescopes. The amplitude calibration was performed following the procedure outlined in \cite{memo}. Phases were then corrected for parallactic angle effects. After an initial correction of instrumental phases and delays using a short scan on a strong calibrator source, fringe-fitting was first performed for all the phase-reference calibrator (J1021+3437, J1157+1638, J2105+0033, and J2303$-$1002) and fringe-finder sources (4C\,39.25, 3C\,454.3). The calibrated data were then exported to the {\sc Difmap} package \citep{difmap} for imaging. The conventional hybrid mapping procedure with subsequent iterations of CLEANing \citep{clean} and phase (then amplitude) self-calibration provided the images and brightness distribution models for the calibrators. Overall antenna gain correction factors were determined at the first step of the amplitude self-calibration. These indicated that the a-priori amplitude calibration was accurate within $5$\,\% for each antenna, therefore no further gain correction was necessary for the target sources. Fringe-fitting was repeated for the phase-reference calibrators in {\sc AIPS}. For this second case, the CLEAN component models of their brightness distributions derived in {\sc Difmap} were also taken into account, to compensate for any residual phases resulting from their non-pointlike structure. The solutions obtained were interpolated and applied to the respective target source data. Finally, the calibrated and phase-referenced visibility data of the four target sources (J1023$+$3243, J1155$+$1507, J2104$-$0009, and J2304$-$0933) were also exported to {\sc Difmap} for imaging. Natural weighting was applied to achieve the lowest image noise. No self-calibration was attempted. The images displayed in Figures~\ref{fig:J1155}--\ref{fig:J2304} were restored with CLEAN component models obtained in {\sc Difmap} for the three target sources detected.  

To describe the brightness distribution quantitatively, we fitted the visibilities with circular Gaussian model components using the task {\sc modelfit} within {\sc Difmap}. The parameter errors were calculated following the equations given by \cite{error}; to account for an assumed VLBI amplitude calibration uncertainty, another $10$\,\% error was added quadratically to the flux density errors. The sum of the integrated flux densities of these model components ($S_\mathrm{VLBA}$) for each source is given in Col. 6 in Table \ref{tab:param}. The parameters and designations of the model components are listed in Table \ref{tab:comp}.

\section{Results} \label{res}

\begin{figure}
\includegraphics[bb=35 135 585 680, clip=, width=\columnwidth]{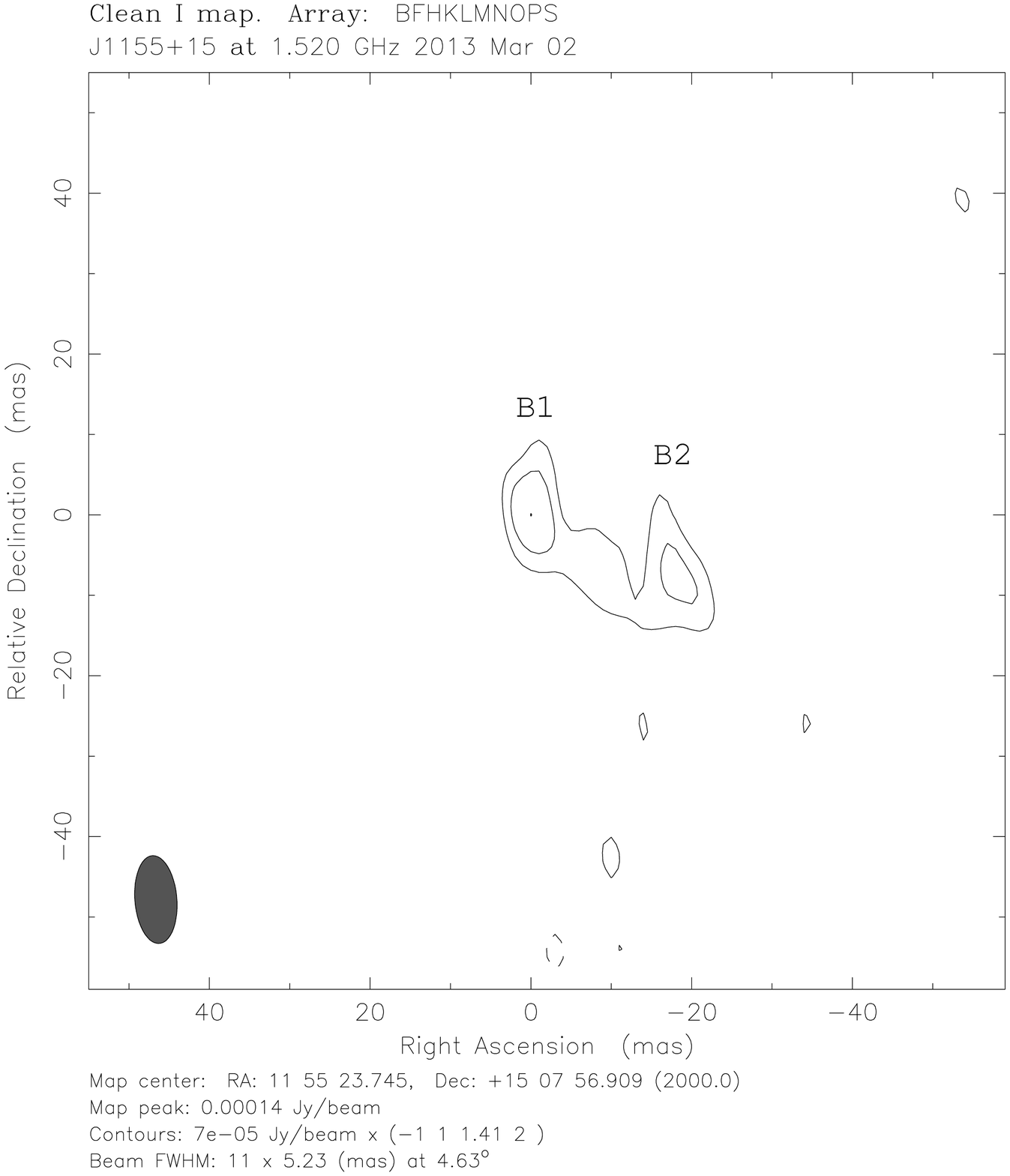}
\caption{\label{fig:J1155}The 1.5-GHz VLBA map of J1155$+$1507. The peak brightness is $140\,\mu$Jy\,beam$^{-1}$. There are three positive contours, the lowest one of those is at $70\,\mu$Jy\,beam$^{-1}$ ($\sim3\sigma$), further contours increase with a factor of $\sqrt 2$. The negative contour (shown by dashed line) is at $-70\,\mu$Jy\,beam$^{-1}$. The restoring beam is $11.0$\,mas\,$\times 5.2$\,mas (FWHM) at a position angle of $4\fdg6$.}
\end{figure}

\begin{figure}
\includegraphics[bb=35 235 585 580, clip=, width=\columnwidth]{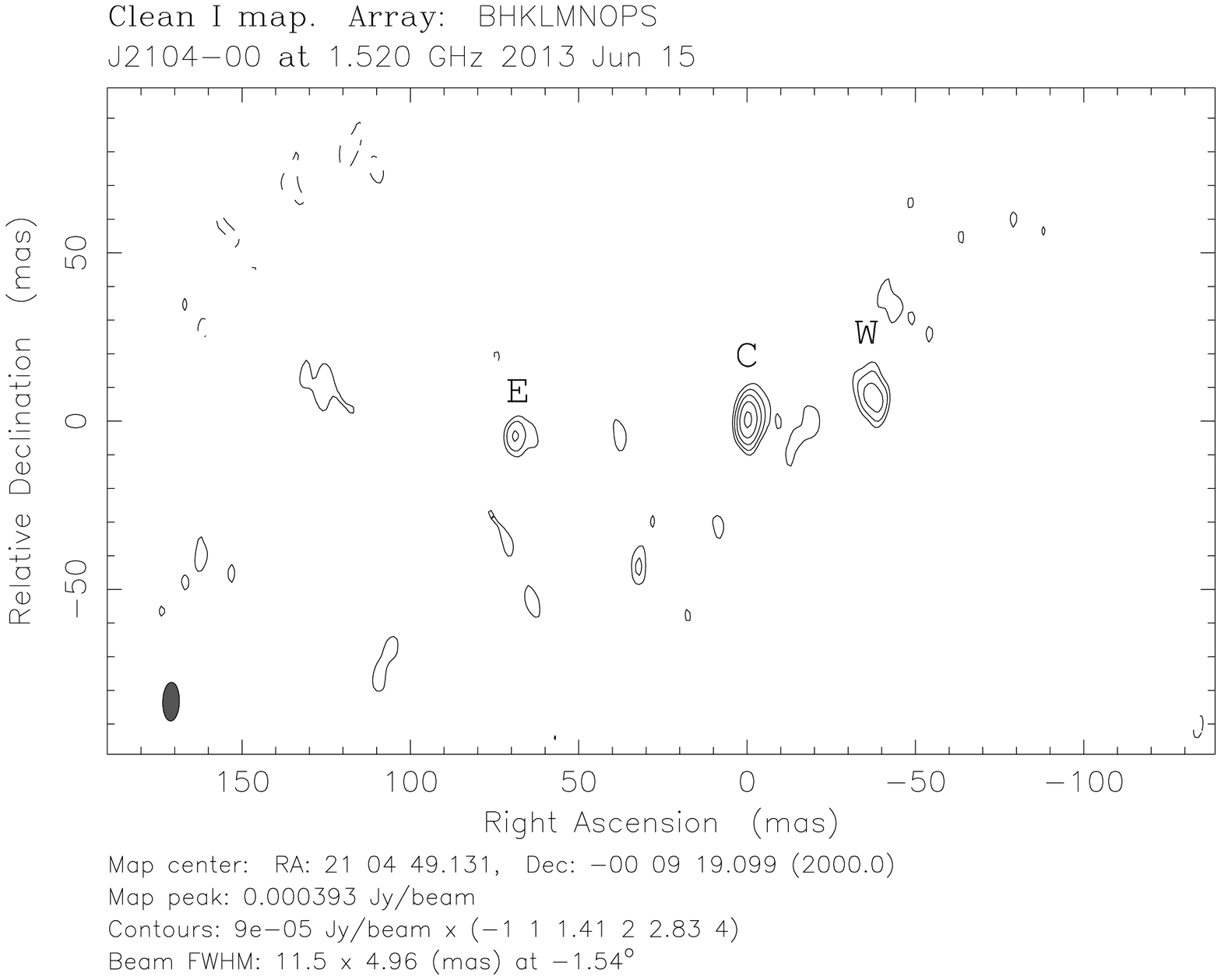}
\caption{\label{fig:J2104}The 1.5-GHz VLBA map of J2104$-$0009. The peak brightness is $393\,\mu$Jy\,beam$^{-1}$. There are five positive contours, the lowest one of those is at $90\,\mu$Jy\,beam$^{-1}$ ($\sim3\sigma$), further contours increase with a factor of $\sqrt 2$. The negative contour (shown by dashed line) is at $-90\,\mu$Jy\,beam$^{-1}$. The restoring beam is $11.5$\,mas\,$\times 5.0$\,mas (FWHM) at a position angle of $1\fdg5$.}
\end{figure}

\begin{figure}
\includegraphics[bb=30 130 768 507, clip=, width=\columnwidth]{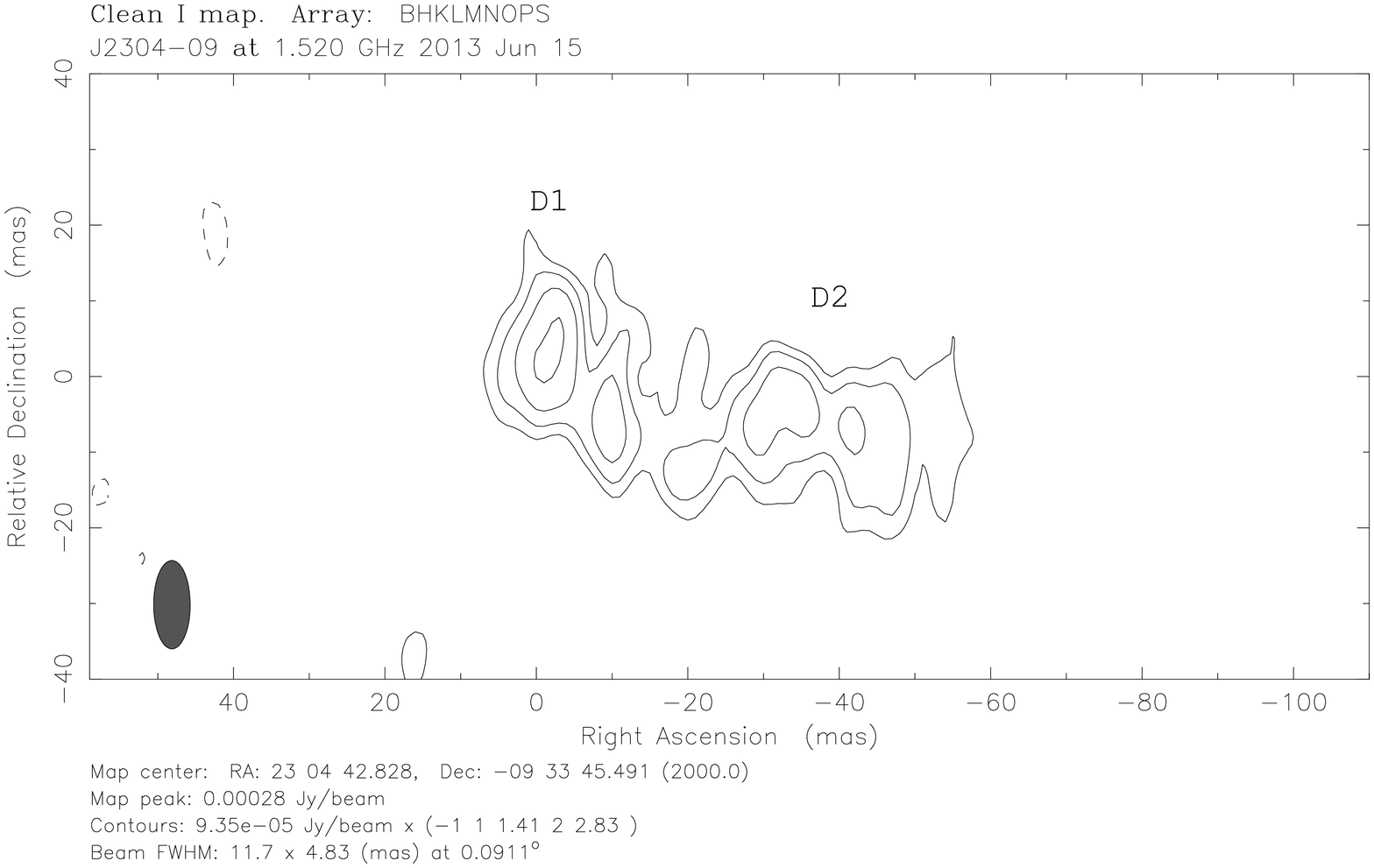}
\caption{\label{fig:J2304} The 1.5-GHz VLBA map of J2304$-$0933. The peak brightness is $280\,\mu$Jy\,beam$^{-1}$. There are four positive contours, the lowest one of those is at $94\,\mu$Jy\,beam$^{-1}$ ($\sim3\sigma$), further contours increase with a factor of $\sqrt 2$. The negative contour (shown by dashed line) is at $-94\,\mu$Jy\,beam$^{-1}$. The restoring beam is $11.7$\,mas\,$\times 4.8$\,mas (FWHM) at a position angle of $0\fdg1$.}
\end{figure}

\begin{deluxetable}{ccccccccc}
\rotate
\tablecaption{\label{tab:param}Parameters of the four candidate dual AGN observed with VLBA.}
\tablewidth{0pt}
\tablehead{
\colhead{Name} & \colhead{RA} & \colhead{DEC} & \colhead{$S_\mathrm{FIRST}$} & \colhead{$P_\mathrm{FIRST}$} &
\colhead{$S_\mathrm{VLBA}$} & \colhead{$P_\mathrm{VLBA}$} & \colhead{$z$} &
\colhead{Scale}  \\
 & & & \colhead{mJy} & \colhead{mJy beam$^{-1}$} & \colhead{mJy} & \colhead{mJy beam$^{-1}$} & & \colhead{pc mas$^{-1}$}
}
\startdata
J1023+3243 & $10^\mathrm{h} 23^\mathrm{m} 25\fs698$\tablenotemark{a} & $+32\arcdeg 43\arcmin 48\farcs47$\tablenotemark{a}  & 2.7 $\pm$ 0.4 & 1.2 $\pm$ 0.1 & \nodata & $< 0.125$ & 0.127 & 2.271\\
J1155+1507 & $11^\mathrm{h} 55^\mathrm{m} 23\fs7449$ & $+15\arcdeg 07\arcmin 56\farcs909$ & 2.0 $\pm$ 0.3 & 1.6 $\pm$ 0.1 & $0.58 \pm 0.16$ & $0.140$ & 0.287 & 4.319 \\
J2104$-$0009 & $21^\mathrm{h} 04^\mathrm{m} 49\fs1306$ & $-00\arcdeg 09\arcmin 19\farcs099$& 3.5 $\pm$ 0.2 & 4.0 $\pm$ 0.1 & $1.37 \pm 0.22$ & $0.393$ & 0.135 & 2.393 \\
J2304$-$0933 & $23^\mathrm{h} 04^\mathrm{m} 42\fs8280$ & $-09\arcdeg 33\arcmin 45\farcs494$ & 8.9 $\pm$ 0.3 &  9.0 $\pm$ 0.2 & $4.37 \pm 0.83$ & $0.280$ & 0.032 & 0.639\\
\enddata

Col. 1: source names; Col. 2: right ascensions; Col. 3: declinations; Col. 4 integrated flux densities from the FIRST catalog \citep{first_new}, the $1\sigma$ errors, since they are not given in the catalog, are derived from the FITS images downloaded from Vizier (http://vizier.u-strasbg.fr/viz-bin/VizieR?-source=VIII/92) ; Col. 5 peak brightness and $1\sigma$ errors from the FIRST catalog \citep{first_new}; Col. 6: sums of the flux density of the Gaussian model components fitted to the VLBA visibilities; Col. 7: peak brightness of the VLBA maps; Col. 8: redshifts; Col. 9: linear scales of the sources.
\tablenotetext{a}{J1023+3243 was undetected with VLBA, thus the coordinates given are from the FIRST survey \citep{first_new}.}
\end{deluxetable}

Three sources were detected out of our four targets. The $5\sigma$ brightness upper limit for the non-detected source, J1023+3243, is $125\mathrm{\, } \mu$Jy\,beam$^{-1}$. According to the FIRST survey, this source has the smallest peak brightness value ($1.2$\,mJy\,beam$^{-1}$) and it is the least compact among the four, indicated by the smallest ratio of the peak brightness to the integral flux density (Table \ref{tab:param}). Therefore it is not surprising that it was undetected in our high-resolution VLBA observation.

We used the {\sc AIPS} verb {\sc maxfit} to derive the astrometric positions of the brightness peaks in the detected three sources. These coordinates do not necessarily coincide with the position of the AGN core. The obtained positions are listed in Col. 2 of Table \ref{tab:param}. We estimate that each coordinate is accurate to within $2$\,milli-arcseconds (mas). The sources of positional error are the thermal noise of the interferometer phases, the error of the phase-reference calibrator positions and the systematic error of the phase-reference observations mainly originating from atmospheric turbulences. In the case of the calibrator source J2303$-$1002, the positional accuracy given in the second realization of the International Celestial Reference Frame \citep[ICRF2,][]{icrf2} is $20$\,mas. However, more accurate coordinates are given in the latest Radio Fundamenal Catalog of L. Petrov (rfc2015c)\footnote{\url{http://astrogeo.org/vlbi/solutions/rfc\_2015c/}}. When we calculated the astrometric position of the target source J2304$-$0933, we used the latter values, thus we were able to obtain the coordinates with the same accuracy as for the other two detected sources.

{\bf J1155$+$1507} is the most distant ($z=0.287$) and the faintest of the three detected sources. It has an elongated structure of $\sim 30$\,mas (Figure \ref{fig:J1155}). The brightness distribution can be best fitted with two circular Gaussian components (Table \ref{tab:comp}), a compact (B1), and a more extended one (B2). Among the three detected sources, this has the largest ratio of the `missing' flux density compared to the FIRST flux density, i.e. more than $70$\,\% of the radio flux density detected on arcsec scale originates from extended emission that is resolved out with the VLBA. 

The radio image of {\bf J2104$-$0009} (Figure \ref{fig:J2104}) is dominated by a compact bright feature (C). It can be well fitted with a circular Gaussian brightness distribution of $(0.60 \pm 0.06)$\,mJy flux density and $(5.0 \pm 0.4)$\,mas diameter (full width at half-maximum, FWHM). Several low-brightness features can be detected at both sides of the compact central component. The elongated structure is rather asymmetrical. Considering the most prominent detected features, it can be traced to $\sim 70$\,mas to the east and $\sim 40$\,mas to the west of component C. These two distant features can be fitted with circular Gaussian components. The eastern one (E) is smaller and fainter, while the western one (W) is slightly more extended and brighter (Table \ref{tab:comp}).

{\bf J2304$-$0933} is the brightest and the closest source ($z=0.032$) in our sample. At the high resolution of the VLBA, it exhibits a complex linear radio structure oriented roughly east--west with a size of almost $70$\,mas (Figure \ref{fig:J2304}). The source seems to be decomposed into 5 or 6 different features along the direction of the minor axis of the restoring beam. However, they cannot be fitted individually, the best fitted model consists of two slightly extended circular Gaussian brightness distribution components, D1 and D2 (Table \ref{tab:comp}). 

\begin{table}
\begin{center}
\caption{Results of the Gaussian model fitting of the three detected sources. \label{tab:comp}}
\begin{tabular}{ccccccc}
\tableline\tableline
Name & Comp. ID & Distance & PA & Flux density & FWHM size & $T_\mathrm{B} $\\
 & & mas & $\arcdeg$ & mJy & mas & $10^6 \mathrm{K}$\\
\tableline
J1155$+$1507 & B1 & & & $0.15 \pm 0.04$ & $3.8 \pm 0.7$ &  $7.1 \pm 4.5 $\\
 & B2 & $16.2 \pm 3.2$ & $ -121 \pm 10$  & $0.43 \pm 0.12$ & $15.9 \pm 4.0$ & $1.2 \pm 0.9$ \\
\hline
J2104$-$0009 & C & & & $0.60 \pm 0.06$ & $5.0 \pm 0.4$ & $14.4 \pm3.7 $ \\
 & E & $68.8  \pm 1.8$ & $94 \pm 1$ & $0.31\pm 0.07$ & $5.7 \pm 0.9$ & $5.7 \pm 3.1$ \\
 & W & $37.3 \pm 0.9 $ & $-78 \pm 1 $ & $0.46 \pm 0.09$ & $7.6 \pm 1.1$ & $4.8 \pm 2.3$ \\
\hline
J2304$-$0933 & D1 & & & $1.70 \pm 0.27$ & $20.7 \pm 2.5$ & $2.2 \pm 0.9  $ \\
  & D2 & $38.0 \pm 4.5 $ & $-105 \pm 6$ & $2.67 \pm 0.56$ & $30.0\pm 4.9$ & $1.6 \pm 0.9$ \\
\tableline
\end{tabular}

Col. 1: source names; Col. 2: component designations; Col 3 and 4: separations and position angles with respect to the brightest component (position angles are measured from north through east); Col. 5: integrated flux densities; Col. 6: FWHM sizes; Col. 7: calculated brightness temperatures.
\end{center}
\end{table}

We calculated the brightness temperature of the Gaussian components in the detected three sources as:
\begin{equation}
T_\mathrm{B}=1.22 \times 10^{12} \frac{S}{\theta^2 \nu^2}(1+z) \mathrm{\,\,K,}
\end{equation}
where $S$ is the flux density measured in Jansky, $\theta$ is the FWHM size of the fitted circular Gaussian component in mas, $z$ is the redshift and $\nu$ is the observing frequency in GHz. The brightness temperatures of each component given in the last column of Table \ref{tab:comp} are typically a few million K, except for component C in J2104$-$0009 which has an order of magnitude higher brightness temperature value, $(14.4 \pm 3.7) \times 10^{6}$\,K. Thus the radio emission from these components are non-thermal, of synchrotron origin. The derived brightness temperature values all exceed the limit of $10^5$\,K, which is the maximum brightness temperature for starburst galaxies not containing an active nucleus \citep{normal_galaxy_tb}. On the other hand, the brightness temperatures are all well below the equipartition brightness temperature \citep[$\sim 5\times 10^{10}$\,K,][]{readhead}, thus there is no indication of relativistic beaming of the radio (jet) emission in any of the sources. 
These values are typical for low luminosity radio AGN \citep[e.g.,][]{Alexandroff2012, LLAGN_VLBI1}. 

We can estimate the monochromatic radio power of the detected VLBI components according to
\begin{equation}
P=4 \pi D_\mathrm{L}^2 S (1+z)^{-\alpha-1}, \label{power}
\end{equation}
where $D_\mathrm{L}$ is the luminosity distance, $S$ is the flux density at $1.5$\,GHz, $z$ is the redshift, and $\alpha$ is the spectral index (defined as $S\sim\nu^{\alpha}$). Compact radio sources usually have flat spectra (thus $\alpha \sim 0$), while more extended, jet-like features have steep spectra with a spectral index typically between $-0.8$ and $-1.1$ \citep[e.g.,][]{Hovatta_spectralindex}. As we do not have spectral information of our sources, for the VLBI-detected compact features we assume a spectral index of $-0.5$. Since the redshifts of the sources are small, the choice of spectral index does not influence significantly the value of the obtained radio powers.
%{\bf Spectral index values larger than this are often associated with a compact source, while smaller values ... with jet-like features. } 
The calculated radio powers are a few times $10^{22}\mathrm{\,W\,Hz}^{-1}$ in the case of the components of J1155$+$1507 and J2104$-$0009, and an order of magnitude lower, $(4-6) \times 10^{21}\mathrm{\,W\,Hz}^{-1}$ in the case of J2304$-$0933. According to \cite{Kewley2000} and \cite{Middelberg2011}, AGN have high-luminosity cores, with power exceeding $\sim 2 \times 10^{21} \mathrm{\,W\,Hz}^{-1}$. On the other hand, \cite{Alexandroff2012} use the luminosities of the supernova remnant (SNR) complexes of Arp 299-A and Arp 220 to determine an upper limit for non-thermal radio emission that can originate from starburst-related activity. These are $P_{1.7\mathrm{\,GHz}} \approx 4.5 \times 10^{21}\mathrm{\,W\,Hz}^{-1}$, and $P_{1.7\mathrm{\,GHz}} \approx 1.6 \times 10^{22}\mathrm{\,W\,Hz}^{-1}$ in the case of Arp 299-A and Arp 220, respectively. The derived radio powers of our faint radio sources are very close to these limiting values, however there are no other indications of starburst activity in any of the objects. All of the sources in \cite{Comerford_sample} were selected using optical emission line diagnostics to ensure the selection of AGN. Additionally, \cite{sfr_data} give star formation rate (SFR) for two of our sources, $0.5\mathrm{\,M}_\odot \mathrm{yr}^{-1}$ for J2104$-$0009, and $2\mathrm{\,M}_\odot \mathrm{yr}^{-1}$ for J2304$-$0933. They classify both sources as AGN as well. These SFR values are much lower than those of the two extreme starburst galaxies mentioned above \citep[$> 100\mathrm{\,M}_\odot \mathrm{yr}^{-1}$, e.g.,][]{arp_SFR}. Thus the non-thermal radio emission detected in our VLBA observations can most probably be explained as originating from the AGN in these sources.

We can also invoke the radio--far-infrared correlation \citep{radio-FIR} to estimate the maximum starburst-related contribution in our sources. None of the sources are listed in the publicly available archives of the $60\mu$m observations, therefore we used the {\it Infrared Astronomical Satellite} (IRAS) Scan Processing and Integration Tool\footnote{Provided by the NASA/IPAC Infrared Science Archive: {\url http://irsa.ipac.caltech.edu/applications/Scanpi/}} (SCANPI) to derive upper limits for the $60\mu$m flux densities from the IRAS archive; this is $\sim 0.1$\,mJy for three sources (J1023$+$3243, J2104$-$0009, and J2304$-$0933). In the case of J1155$+$1507, there is a $4\sigma$ detection according to SCANPI with a $60\mu$m flux density of $S_{60}=0.17\pm0.04$\,mJy. The radio flux densities reported by FIRST in all sources are in excess of the radio emissions calculated from the radio--far-infrared correlation using the upper limit of $0.1$\,mJy and using the $0.17$\,mJy detection. This also indicates that AGN must contribute to the radio emission. 

The radio structures of the detected three sources are compact with sizes well below $1$\,kpc. Following the description of Compact Symmetric Objects (CSOs) as given in \cite{Fanti_cso} (not necessarily symmetric but two-sided radio structure with respect to a weak or undetected core), J1155$+$1507 and J2304$-$0933 can be categorized as CSOs. Their low radio powers and small sizes would place them in the lower left part of the radio power--linear size diagram shown in Figure 1. of \cite{CSO_Tao}. On the other hand, the radio morphology of J2104$-$0009 (e.g., its prominent central component, C) is not typical of CSOs.

The flux densities obtained with the VLBI observations on angular scales of $\sim 10$\,mas are significantly below the flux densities reported in the FIRST survey at a similar observing frequency. The missing flux densities are $70$\,\%, $61$\,\%, $50$\,\%, and $>95$\,\% for the sources, J1155$+$1507, J2104$-$0009, J2304$-$0933, and J1023$+$3243 respectively. This can arise from the different angular resolution (our high-resolution VLBA observations resolve out the large-scale structure in the sources), or from flux density variability, or the combination of both effects. While flux density variability cannot be ruled out completely, we expect that compact features (on VLBI scales) are responsible for the variability. It is however unlikely that in all sources the most compact features decreased in flux density by a factor of $60$\,\% to $95$\,\% between their observations in the FIRST survey and our VLBA observations. On the other hand, in all detected sources we see complex, resolved VLBI structures. Thus, it is natural to expect that (at least most of) the missing flux densities belong to extended features larger than the largest recoverable size ($\theta_\mathrm{LAS}$) in our VLBA observations. The largest recoverable size of an interferometer is set by its shortest baseline ($B_\mathrm{min}$) and the observing wavelength ($\lambda$) as $\theta_\mathrm{LAS}\sim \lambda/(2\cdot B_\mathrm{min})$ \citep{Wrobel_LAS}. In our VLBA observation, the shortest baseline length is that of between Los Alamos and Pie Town ($236$\,km), thus the array at $1.5$\,GHz is insensitive for structures larger than $\sim 90$\,mas.

We can estimate the radio power connected to these missing flux densities; these are thus upper limits of the radio powers originating from the extended structures of the sources. The values are $\sim 10^{23}\mathrm{\,W\,Hz}^{-1}$ in J1155$+$1507, J2104$-$0009, and J1023$+$3243 and an order of magnitude lower for J2304$-$0933. Some of the radio emission might originate from starburst activity in the AGN host galaxy. Using the relation of \cite{sfr_radiopower}, one can estimate the star formation rate from the radio power as:
\begin{equation}
SFR = \frac{P_{1.4\mathrm{\,GHz}}}{1.8 \times 10^{21} \mathrm{W\,Hz}^{-1}} \mathrm{\,M}_\odot \mathrm{yr}^{-1}
\end{equation}
Knowing the upper limit for the radio power, we can derive an upper limit for the star formation rate. These are $18\mathrm{\,M}_\odot \mathrm{yr}^{-1}$, $55\mathrm{\,M}_\odot \mathrm{yr}^{-1}$, $6\mathrm{\,M}_\odot \mathrm{yr}^{-1}$, and $61\mathrm{\,M}_\odot \mathrm{yr}^{-1}$ for J1155$+$1507, J2104$-$0009, J2304$-$0933, and J1023$+$3242 respectively. In the two sources for which SFR values are given in the literature \citep[J2104$-$0009, J2304$-$0933; ][]{sfr_data}, the upper limits are consistent with those, thus part of the missing flux density can originate from star formation, but AGN related radio emission (from e.g., extended lobes) plays an important role.

\section{Discussion}

J1155$+$1507 was one of the $3$ candidate dual AGN selected by \cite{Comerford_sample} based on the adaptive optics images of \cite{earlyFu} \citep[see also ][]{1155_Fu_NIRimage}, while J2104$-$0009, J2304$-$0933, and J1023$+$3243 were selected on the basis that their emission features aligned with the major axis of the host galaxy. \cite{Comerford_sample} determined the spatial extent and position angle of the two emission components in their sample. The angular projected spatial offsets of the narrow-line emission features are $0\farcs55$, $0\farcs28$, and $0\farcs84$ in the case of J1155$+$1507, J2104$-$0009, and J2304$-$0933, respectively. These separations are an order of magnitude (in the case of J1155$+$1507, and J2304$-$0933), or more than $2.5$ times (in the case of J2104$-$0009) larger than the size of the radio structures detected in our VLBA observations. 

We checked the $4\arcsec \times 4\arcsec$ fields around the four sources' positions, but we did not find any additional radio emission above $\sim 5\sigma$ image brightness level, thus above $117\,\mu$Jy\,beam$^{-1}$, $150\,\mu$Jy\,beam$^{-1}$, and $157\,\mu$Jy\,beam$^{-1}$ in the case of J1155$+$1507, J2104$-$0009, and J2304$-$0933, respectively. Using equation \ref{power}, and assuming a spectral index, we can estimate the monochromatic the radio power upper limits associated with these image brightness levels. The largest (most conservative) upper limits are obtained assuming a steep spectral index of $-1$. The estimated upper limits for the monochromatic powers, $3 \times 10^{22}$\,W\,Hz$^{-1}$ in the case of J1155$+$1507, and $7 \times 10^{21}$\,W\,Hz$^{-1}$ in the case of J2104$-$0009, do not exclude the existence of a low-luminosity radio-emitting AGN. In the case of the third source, J2304$-$0933, the derived upper limit of the radio power is $4 \times 10^{20}$\,W\,Hz$^{-1}$, therefore considering the lower limit of radio power for AGN cores ($2 \times 10^{21}$\,W\,Hz$^{-1}$) given by \cite{Kewley2000} and \cite{Middelberg2011}, we can exclude the existence of another radio-emitting AGN around this source. However a radio-silent AGN can still give rise to the optical emission features detected by \cite{Comerford_sample} and remain undetected in our observation.

We can compare the position angles of the optical emission features and the radio structure as well. Our fitted Gaussian model components to the visibility data, W and E in J2104$-$0009 are at position angles $-78\arcdeg \pm 1\arcdeg$ and at $94\arcdeg \pm 1\arcdeg$, respectively. This roughly east--west oriented structure (Figure \ref{fig:J2104}) matches well with the position angle of $98\arcdeg \pm 5\arcdeg$ of the optical emission components given by \cite{Comerford_sample}. In J2304$-$0933 (Figure \ref{fig:J2304}) the position angle of D2 is $-105\arcdeg \pm 6\arcdeg$ which is broadly consistent (modulo $180\degr$) with the position angle of $66\arcdeg \pm 3\arcdeg$ given by \cite{Comerford_sample}. Thus in these two sources the VLBI-detected radio features, similarly to the optical emission features, are located along the plane of the host galaxy. Radio emission aligned with the plane of the host galaxy might indicate that it originates from star formation rather than AGN jet activity. However, in Sect. \ref{res}, we showed that the high radio powers in J2104$-$0009 and the star formation rate estimate for J2304$-$0933 render unlikely that these radio emission features originate from SNRs or SNR complexes. 

J2104$-$0009 and J2304$-$0933 were imaged in the near-infrared with the NIRC2 camera (PI: Keith Matthews) enhanced by Laser Guide Star Adaptive Optics system on the Keck II Telescope \citep[Keck II/LGSAO, ][]{Keck_LSGAO}. J2104$-$0009 was observed with a broadband {\it H}($1.337-1.929\mu$m) filter, J2304$-$0933 was observed with a broadband {\it Kp}($1.948-2.299\mu$m) filter. The images were published by \cite{McGurk}. Both sources show single spatial structures. \cite{McGurk} concluded that the double-peaked narrow-line emissions in these sources are likely not caused by kpc-separation dual AGN but are due to NLR kinematics, outflows, jets, rings of star formation, or a close pair of AGN which cannot be resolved in the near-infrared. The good coincidence between the double emission components and the radio structure in our VLBA images (Figs. \ref{fig:J2104}, \ref{fig:J2304}) suggest that the radio jets, although on much larger scales than mapped here, do play a role in producing the emission regions responsible for the double-peaked spectral lines. Thus our data support the jet-driven scenario for the double-peaked emission lines. Both of these sources were selected by \cite{Comerford_sample} as good candidates of being dual AGN, because of the orientation of the optical emission features in the plane of the host galaxy. Our observations together with the results of near-infrared imaging seems to suggest that this criterion may not be a good indicator of AGN duality.

For J1155$+$1507, the position angle of the two emission components given by \cite{Comerford_sample} is $137\arcdeg \pm 3\arcdeg$. The position angle of the radio structure (Figure \ref{fig:J1155}) is $-121\arcdeg \pm 10\arcdeg$, nearly perpendicular the position angle derived from optical spectroscopy, thus they are probably not related.
According to the high-resolution image of J1155$+$1507 obtained with the NIRC2 camera on Keck II/LGSAO with a broadband {\it Kp}($1.948-2.299\mu$m) filter by \cite{earlyFu} \citep[see also][]{1155_Fu_NIRimage}, this source is a merging system with a component separation of $2.5$\,kpc. The component separation and the position angle are also compatible with the values given in \cite{Comerford_sample}. Additionally, integral-field spectroscopy of the source with the Supernova Integral-Field Spectrograph \citep{SNIFS1, SNIFS2} on the University of Hawaii $2.2$\,m telescope on Mauna Kea was acquired by \cite{1155_Fu_NIRimage}.
The integral-field spectroscopy data spatially resolved the emission line in the source; \cite{1155_Fu_NIRimage} classified J1155$+$1507 as having extended NLR. The resolved stellar components were enveloped by the spatially extended [\ion{O}{3}] emission. \cite{1155_Fu_NIRimage} concluded that although the extended NLR is clearly responsible for the double-peaked line profiles, it is impossible to decide whether both or just one AGN powers the NLR. In our VLBA image, the detected two components are separated by $\sim 70$\,pc, and connected with continuous emission structure; their radio emission most probably originate in one of the two merging galaxies detected in the near-infrared image \citep{1155_Fu_NIRimage}. Because of the lack of astrometric registration of the adaptive-optics image \citep{1155_Fu_NIRimage}, we cannot determine which near-infrared component is associated with the VLBI-detected compact radio emission. 

The fourth source in our sample, {\bf J1023$+$3243}, which was not detected with VLBA in our experiment, was however successfully observed and imaged by \cite{Muller-Sanchez} with the VLA\footnote{Unfortunately \cite{Muller-Sanchez} do not list the date of their observations. According to the NRAO archive, within their program, the observations of J1023$+$3243 took place on 2012 October 13 and 2013 January 5. Both of these dates precede by several months our VLBA observation (2013 March 2).}. \cite{Muller-Sanchez} observed several sources of the sample of \cite{Comerford_sample} with the VLA in A configuration in X band to clarify the nature of the of the double-peaked narrow line emission galaxies. The only overlapping source with our sample is J1023$+$3243.
\cite{Muller-Sanchez} used two subbands within the X band, the lower one centered at $8.5$\,GHz, the upper one at $11.5$\,GHz. They detected two radio components in J1023$+$3243 when combining the observations from the two subbands by averaging across all spectral channels. The position angle of the radio structure is in agreement within the errors with the results given by \cite{Comerford_sample} from the optical spectroscopy. The position angle agrees also well with the photometric major axis of the host galaxy. The separation of the two radio features are slightly smaller than that of the optical emission line features. The two radio features are also detected in the lower subband, while the detection of the fainter component is only marginal in the upper subband; the fainter feature is detected at $3.4\sigma$ noise level at $11.5$\,GHz.
The VLA-detected radio features are weak with flux densities of $(0.03-0.25)$\,mJy. The spectrum of the brighter feature is slightly inverted, $\alpha=0.25 \pm 0.16$, while the fainter feature is flat with $\alpha=-0.29 \pm 0.19$. (However, because of the marginal detection of the latter feature in the upper subband, the spectrum can be much steeper.) \cite{Muller-Sanchez} conclude that the VLA observation confirm the existence of dual AGN in J1023$+$3243, however based upon the difference between the distance derived from the optical emission and the distance between the VLA-detected radio features, they suggest that additional kinematic component related to outflows may be present in the source.

To interpret the VLA results of  \cite{Muller-Sanchez} and our own VLBA observations of J1023$+$3243 in a common context, we assume that the radio spectra of the two features follow the same power-law down to $1.5$\,GHz. Then the two VLA components have a flux density of $0.16$\,mJy and $0.1$\,mJy, respectively, at 1.5\,GHz. Our VLBA observation would have had adequate sensitivity to detect the brighter one. Our non-detection may be explained by source variability, or resolved source structure (which is already hinted by the very low value, less than $0.5$ of the ratio of peak brightness to integrated flux density in the FIRST, see Table \ref{tab:param}), or an even more inverted spectrum ($\alpha>0.4$) than the one derived from the X-band VLA data. 

At $1.4$\,GHz (in L band), the source was detected in the FIRST survey with a flux density of $2.7$\,mJy, but remained undetected in the NRAO VLA Sky Survey \citep[NVSS,][]{nvss}. The latter has a completeness level of $\sim 2.5$\,mJy, thus it may have just missed J1023$+$3243. (Since the NVSS was conducted at lower resolution than the FIRST, the non-detection cannot be explained by resolution effect.)
One can also compare the extrapolated (to $1.4$\,GHz) and summed flux density of the features detected by \cite{Muller-Sanchez}, $0.26$\,mJy, to the value given in FIRST. If the assumption about the power-law spectra with the X-band spectral indices holds, only one-tenth of the FIRST flux density is detected by \cite{Muller-Sanchez}. This, together with the very low value (less than 50\,\%) of the ratio of peak brightness to integrated flux density in the FIRST, already suggests a resolved radio structure.

Thus we cannot strengthen the conclusion of \cite{Muller-Sanchez} that J1023$+$3243 is a dual AGN, since we did not detect any compact (on VLBA scales), high radio power feature in this source. If there is a compact radio-emitting core in any of the two features detected in the VLA observations, they must have low radio luminosity, a few times $10^{21}\mathrm{\,W\,Hz}^{-1}$, similar to J2304$-$0933. Interestingly, dual AGN scenario for J1023$+$3243 was not supported by the near-infrared image of \cite{McGurk}, since only a single spatial structure was detected.

\section{Summary}

According to hierarchical structure formation models, we expect to see merging galaxies in different stages of their coalescence. Dual AGN with kpc-scale separation bridge the gap between the earliest (galaxy pairs at tens of kpc separations) and ultimate (SMBH binaries at pc or sub-pc separations) evolutionary stages of galaxy mergers, thus they provide vital information on the galaxy evolution and SMBH growth during this intermediate stage \citep{BBHEvo, Komossa_obssum}. 

Because high spatial resolution is required, it is difficult to directly map dual AGN, especially in the distant Universe. Instead, the search for spatially unresolved dual AGN relies mostly on indirect evidence. However, currently there is no known observational signature which can be used to reliably select dual AGN candidates. Double-peaked narrow emission lines in which the radial velocity offset between two line features marks the orbital motion of dual AGNs were proposed as such \citep{wang2009}. But further observations of these candidate sources with high resolution found that only 1\,\% of them can be genuine dual AGN \citep{earlyFu}; in most of the sources this spectral behaviour can be be explained by other effects not requiring dual AGN \citep{Heckman1981, Heckman1984, double_line2, non_dual}. As shown by several works \citep[e.g.,][]{earlyFu, Crenshaw2, Shen_binaryAGN, Tingay_VLBI, Frey_nondual, Comerford_dualsearch2, 3C316_Tao, ngc5515, Kharb, McGurk}, double-peaked narrow lines alone are not good indicators to select compelling dual AGN candidates, other diagnostic methods have to be used.

\cite{Comerford_sample} used high-resolution long-slit spectroscopy to select the most compelling dual AGN
candidates from the double-peaked narrow emission line active galaxies. They concluded that $17$ out of the observed $84$ are good dual AGN candidates. We observed with the VLBA at $1.5$\,GHz four of the $17$ candidate sources, which have the highest radio flux densities according to the FIRST survey. These observations were intended as a pilot study aimed to those sources where results can be obtained with reasonable integration time. Thus our sample is not complete. To obtain a statistically useful result more observations of the fainter sources are essential. 

From the observed sources, we were able to detect three. None of them has a secondary compact radio-emitting source at a separation indicated by the optical emission line features \citep{Comerford_sample} at a $5\sigma$ level. In the case of J2304$-$0933, the lower limit of the radio power calculated from the rms noise level is less than the lower limit implied for the cores of radio-emitting AGN by \cite{Kewley2000} and \cite{Middelberg2011}. Thus in the case of J2304$-$0933, we can exclude the existence of another radio-emitting AGN in the host galaxy. In the other two detected sources, low-luminosity ($P_\mathrm{1.5\,GHz}<3\times10^{22}$\,W\,Hz$^{-1}$ and $P_\mathrm{1.5\,GHz}<7\times10^{21}$\,W\,Hz$^{-1}$) radio-emitting AGN can in princple be still present in the sources.

In two sources (J2104$-$0009 and J2304$-$0933), the radio structures are oriented at similar position angles as the optical emission features. The radio components in these two sources may be associated with jet emission, thus the double-peaked emission lines can be caused by jet-driven outflows, instead of dual AGN. However, in the third source (J1155$+$1507), where \cite{1155_Fu_NIRimage} imaged and spatially resolved the two interacting galaxies in the near-infrared positionally coinciding with the double-peaked emission line features, the observed VLBA radio structure is less extended and is oriented in a nearly perpendicular position angle. Thus, the radio structure is seemingly unrelated to the structure seen in the narrow emission line emitting regions. Likely one of the merging galaxies has a radio-emitting AGN, but we cannot determine which galaxy is associated with the compact radio emission.

The fourth source (J1023$+$3243) which remained undetected in our VLBA mini-survey, was however detected by \cite{Muller-Sanchez} with the VLA at $8.5$\,GHz and  $11.5$\,GHz a few months before our observations. They observed two radio components at a position angle corresponding to that of the optical emission features. Using the derived spectral index of the brighter feature, if the source is not variable, and/or if it is compact, we should have been able to detect it with the VLBA. FIRST survey data suggest that the source has extended, resolved structure at $1.4$\,GHz. 
Thus we cannot strengthen the conclusion of \cite{Muller-Sanchez} that J1023$+$3243 is a dual AGN. 

In summary, we did not find two compact radio emitting cores in any of the four VLBA-observed candidate dual-AGN sources. This does not exclude the presence of dual AGN in these sources, since $\sim 90$\,\% of AGN are known to be radio-quiet. %But even if they are dual AGN then at most only one nucleus is active in radio. 

VLBI observations are an effective way to prove AGN-related radio emission and can provide the required resolution to confirm the existence of the two radio-emitting nuclei in dual AGN candidates \citep[e.g.,][]{triple_Deane}. However, the majority of the AGN are not radio-loud and currently it is not clear whether the triggering of radio emission is related anyhow to duality or merging phases. On the other hand, mapping the jet structure with radio observations and compare it to structures associated with those emitting the optical spectral features might indicate that the double-peaked emission lines originate from jet-driven outflows. For nearby sources, mapping the spectral index distribution at VLA resolution may help in this regard. For example, \cite{radioselected_binaryAGN2} discarded two of their six candidate dual AGN sources \citep{radioselected_binaryAGN} based upon the radio spectral index maps. But to reliably distinguish jets and outflows from the AGN core, the fine resolution provided by the VLBI technique is important.

Achieving the required imaging sensitivity to completely rule out the existence of two low-luminosity radio-emitting AGN in a candidate source requires ample amount of observing time (preferably at more than one frequency), thus with currently available instruments it can realistically be accomplished for individual sources or small samples, and not in survey mode. The Square Kilometre Array (SKA), when fully deployed would have the powerful survey efficiency and the sufficient sensitivity to detect dual AGN with separations less than $1$\,kpc in principle at all redshifts \citep{Deane_SKA}. The SKA as an element of a VLBI array \citep{SKA-VLBI} could be necessary to identify close pairs at pc-scale separations.

\acknowledgments
We thank the anonymous referee for helpful comments.
The National Radio Astronomy Observatory is a facility of the National Science Foundation operated under cooperative agreement by Associated Universities, Inc. This work made use of the Swinburne University of Technology software correlator, developed as part of the Australian Major National Research Facilities Programme and operated under licence \citep{VLBA_softcorr}. This research has made use of the NASA/IPAC Infrared Science Archive, which is operated by the Jet Propulsion Laboratory, California Institute of Technology, under contract with the National Aeronautics and Space Administration.
This work was supported by the China Ministry of Science and Technology 973 programme under grant No. 2013CB837900, the Hungarian National Research, Development and Innovation Office (OTKA K104539, NN110333), and the China--Hungary Collaboration and Exchange Programme by the International Cooperation Bureau of the Chinese Academy of Sciences (CAS). This research has made use of the NASA/IPAC Extragalactic Database (NED) which is operated by the Jet Propulsion Laboratory, California Institute of Technology, under contract with the National Aeronautics and Space Administration. 

%% To help institutions obtain information on the effectiveness of their
%% telescopes, the AAS Journals has created a group of keywords for telescope
%% facilities. A common set of keywords will make these types of searches
%% significantly easier and more accurate. In addition, they will also be
%% useful in linking papers together which utilize the same telescopes
%% within the framework of the National Virtual Observatory.
%% See the AASTeX Web site at http://aastex.aas.org/
%% for information on obtaining the facility keywords.

%% After the acknowledgments section, use the following syntax and the
%% \facility{} macro to list the keywords of facilities used in the research
%% for the paper.  Each keyword will be checked against the master list during
%% copy editing.  Individual instruments or configurations can be provided 
%% in parentheses, after the keyword, but they will not be verified.

{\it Facilities:} \facility{NRAO VLBA}
%, \facility{HST (STIS)}, \facility{CXO (ASIS)}.

%% The reference list follows the main body and any appendices.
%% Use LaTeX's thebibliography environment to mark up your reference list.
%% Note \begin{thebibliography} is followed by an empty set of
%% curly braces.  If you forget this, LaTeX will generate the error
%% "Perhaps a missing \item?".
%%
%% thebibliography produces citations in the text using \bibitem-\cite
%% cross-referencing. Each reference is preceded by a
%% \bibitem command that defines in curly braces the KEY that corresponds
%% to the KEY in the \cite commands (see the first section above).
%% Make sure that you provide a unique KEY for every \bibitem or else the
%% paper will not LaTeX. The square brackets should contain
%% the citation text that LaTeX will insert in
%% place of the \cite commands.

%% We have used macros to produce journal name abbreviations.
%% AASTeX provides a number of these for the more frequently-cited journals.
%% See the Author Guide for a list of them.

%% Note that the style of the \bibitem labels (in []) is slightly
%% different from previous examples.  The natbib system solves a host
%% of citation expression problems, but it is necessary to clearly
%% delimit the year from the author name used in the citation.
%% See the natbib documentation for more details and options.

\bibliography{dual_ref}

\end{document}